# Lump Solutions to a Jimbo-Miwa Like Equations

Harun-Or-Roshid[a,*], M. Zulfikar Ali[b]

[a]Department of Mathematics, Pabna University of Science and Technology, Bangladesh
[b]Department of Mathematics, Rajshahi University, Bangladesh
[*]Email: harunorroshidmd@gmail.com

**Abstract**

A Jimbo-Miwa like nonlinear differential equation in (3+1)-dimensions is developed through a generalized bilinear equation with the generalized bilinear derivatives $D_{3,x}$, $D_{3,y}$, $D_{3,z}$ and $D_{3,t}$. Based on the generalized bilinear forms, two classes of lump solutions, rationally localized in all directions in the space, and a class of complex lump type solution are generated from a Maple search of quadratic polynomial function solution to the proposed Jimbo-Miwa like equation. The apposite conditions to assurance analyticity and rational localization of the solutions are offered. Each of the resulting lump solutions hold six free parameters, two of which are due to the translation invariance of the Jimbo-Miwa like equations and four of which only require to satisfy the presented apposite conditions of being lump solutions. We also generate circle 'O' or ellipse shape solutions from the obtained solutions using different conditions on the lumps solutions. Moreover, figures are given to visualize the properties of the explicit solutions.

## 1. Introduction

Soliton equations connect rich histories of exactly solvable systems constructed in mathematics, fluid physics, microphysics, cosmology, field theory, etc. To explain some physical phenomenon further, it becomes more and more important to seek exact rational solutions, especially rogue wave solutions. Recently, there has been increasing interest in finding rational solutions, lump a kind of rogue wave like solution, to nonlinear differential equations. A special kind of rational solutions – rogue wave solutions – draws a major awareness of researchers worldwide, which illustrate significant nonlinear wave phenomena in oceanography [1,2] and nonlinear optics [3,4]. Rational solutions of the integrable differential equations have been steadily considered with the help of the Wronskian formulation or the Casoratian formulation (see, e.g., [5–7]) based on Hirota bilinear forms. One of recent interests to us is to discuss about rational solutions to a novel class of nonlinear differential equations related with bilinear equations. Real valued solutions by means of rationally decay or lumps have been broadly deliberated in recent researches. Lumps (in the Refs.[8] and [9]) for the DS-II and the Sine-Gordon Equations are established. Nontrivial dynamics of lumps to the KPI [10] are deliberated in the literature. This solutions exhibits interesting scattering properties that were first noticed in Ref.[11]. The Jimbo-Miwa [12-15] equation is used to describe certain interesting (3+1)-dimensional waves in physics and is the second equation in the well known Painlev`e hierarchy of integrable systems. The simplest non-linear (3+1)-dimensional Jimbo-Miwa equation [12-15]:

$$u_{xxxy} + 3u_{xx}u_y + 3u_{yx}u_x + 2u_{yt} - 3u_{xz} = 0, \tag{1}$$

Many researches expend considerable efforts to solve the Eq.(1). Tang and Liang [12] studied the equation through the multi-linear variable separation scheme. Xu [13] tested the integrability through the Painlevé test and showed that Eq. (1) is not integrable and through the obtained truncated Painlevé expansions established two bilinear equations to analyze one soliton, two soliton and dromin solutions. Dai et. al. [14] used two-soliton method, bilinear method and transforming parameters into complex ones method to achieve a new class of cross kink-wave and periodic solitary-wave solution to the Eq.(1). Wazwaz [15] employed Hirota's bilinear method to develop mult-soliton solutions for the Eq.(1). Asaad and Ma [16] applied bilinear Bäcklund transforms to find extended gram-type determinant and find wave, rational solutions to the (3+1)-dimensional Jimbo-Miwa equation.

In this article, we would like to focus on the (3+1)-dimensional Jimbo-Miwa like equation from its bilinear form and would like to present two classes of lump solutions and a class of lump type complex solutions reducing dimension of the equation. In this present paper, we show that few sufficient conditions should be satisfied for the solutions being lump type solutions. Otherwise, they give different shaped solutions even singular solutions. All the solutions are illustrated with 3D plot and density plot.

## 2. Formulation of the Jimbo-Miwa like Equation and its Bilinear forms

In this section, we would like to construct a Jimbo-Miwa like equation. To do that at first transforming the (3+1)-dimensional nonlinear Jimbo-Miwa equation Eq. (1) into the bilinear forms through the dependent variable transformations:

$$u = 2(\ln f)_x. \tag{2}$$

The above (3+1)-dimensional nonlinear evolution Eq. (1) is mapped into the Hirota bilinear equations: $(D_x^3 D_y + 2D_y D_t - 3D_x D_z)f \cdot f = 0,$ (3)

where the bilinear differential equation is defined as

$$\prod_{i=1}^{M} D_{x_i}^{n_i} f \cdot g = \prod_{i=1}^{M} \left(\frac{\partial}{\partial x_i} - \frac{\partial}{\partial x'_i}\right)^{n_i} f(x)g(x')\bigg|_{x'=x}, \tag{4}$$

where $x = (x_1, \cdots, x_M)$, $x' = (x'_1, \cdots, x'_M)$ and $n_1, \cdots, n_M$ are arbitrary nonnegative integers. More precisely, under the virtue of the Eq. (2), the Eq.(3) is reduced to,

$$f(f_{xxxy} + 2f_{yt} - 3f_{xz}) - f_{xxx}f_y - 3f_{xxy}f_x + 3f_{xx}f_{xy} - 2f_y f_t + 3f_x f_z = 0. \tag{5}$$

We would like to consider another bilinear differential equation, a generalized bilinear differential equation similar to the Jimbo-Miwa equation as

$$(D_{3,x}^3 D_{3,y} + 2D_{3,y}D_{3,t} - 3D_{3,x}D_{3,z})f \cdot f = 0. \tag{6}$$

This is the same bilinear form as in the basic Jimbo-Miwa equation Eq.(3). The operators adopted beyond are a kind of generalized bilinear differential operators linked with prime number $p = 3$, which are known as in various article [17, 18]:

$$\prod_{i=1}^{M} D_{p,x_i}^{n_i} f \cdot g = \prod_{i=1}^{M} \left(\frac{\partial}{\partial x_i} + \alpha \frac{\partial}{\partial x_i'}\right)^{n_i} f(x) g(x') |_{x'=x}, \quad (7)$$

where $x = (x_1, \cdots\cdots, x_M)$, $x' = (x_1', \cdots\cdots, x_M')$, $n_1, \cdots, n_M$ are arbitrary nonnegative integers, and $m$ th power of $\alpha$ is defined by

$$\alpha^m = (-1)^{r(m)}, \text{if } m \equiv r(m) \bmod p, \ 0 \le r(m) < p.$$

If $p = 3$, in particular we have

$$\alpha = -1, \alpha^2 = \alpha^3 = 1, \alpha^4 = -1, \alpha^5 = \alpha^6 = 1, \cdots\cdots,$$

Thus, when $p = 3$, we have

$$D_{3,x}^3 D_{3,y} f \cdot f = 6 f_{xx} f_{xy}, \ D_{3,y} D_{3,t} f \cdot f = 2 f_{ty} f - 2 f_t f_y, \ D_{3,x} D_{3,z} f \cdot f = 2 f_{xz} f - 2 f_x f_z. \quad (8)$$

More precisely, under the virtue of the Eq. (2), the Eq.(6) is reduced to,

$$2(3 f_{xx} f_{xy} + 2 f_{ty} f - 2 f_t f_y - 3 f_{xz} f + 3 f_x f_z) = 0. \quad (9)$$

According to a general bell polynomial theories (see, e.g., [17, 18]), we adopt a dependent variable transformation Eq.(2) to transform bilinear equations to nonlinear equations. Then it can directly be shown that the generalized bilinear Eq.(6) or Eq.(9) is linked to a Jimbo-Miwa like nonlinear differential equation:

$$\begin{cases} \frac{9}{8} u^2 u_x v + \frac{3}{8} u^3 u_y + \frac{3}{4} uvu_{xx} + \frac{3}{4} u_x^2 v + \frac{3}{4} u^2 u_{xy} + \frac{9}{4} uu_x u_y + \frac{3}{2} u_{xx} u_y + \frac{3}{2} u_x u_{xy} + 2 u_{yt} - 3 u_{xz} = 0 \\ u_y = v_x. \end{cases} \quad (10)$$

Actually under the Eq.(2), we have the following equality:

$$\left[\frac{(D_{3,x}^3 D_{3,y} + 2 D_{3,y} D_{3,t} - 3 D_{3,x} D_{3,z}) f \cdot f}{f^2}\right]_x = \frac{9}{8} u^2 u_x v + \frac{3}{8} u^3 u_y + \frac{3}{4} uvu_{xx} + \frac{3}{4} u_x^2 v + \frac{3}{4} u^2 u_{xy}$$
$$+ \frac{9}{4} uu_x u_y + \frac{3}{2} u_{xx} u_y + \frac{3}{2} u_x u_{xy} + 2 u_{yt} - 3 u_{xz} = 0, \quad (11)$$

where $u_y = v_x$.

Since $u = 2(\ln f)_2$, is translates PDEs to its bilinear form. Therefore, if $f$ solves Eq.(9), then transformation $u = 2(\ln f)_x$ will solve the Jimbo-Miwa like equation Eq.(10). The Eq.(10) has more extra terms with higher nonlinearity than the usual Jimbo-Miwa equation. Beside this, in contrast their bilinear counterparts, we scrutinize a different phenomenon that the generalized bilinear Jimbo-Miwa Eq.(9) or Eq.(10) is much simpler than that of the usual Jimbo-Miwa Eq.(5) or Eq.(1).

### 3. Lump Solutions of the Jimbo-Miwa like Equations:

With the aid of the symbolic computational software Maple/mathematica, we are going to inquire for positive quadratic solutions to the bilinear equations reducing dimension via $z = x$ or $z = y$ in Eq (9). In the two-dimensional space, a solution involved summing of one square does not generate exact solutions which are rationally localized in all directions in the space, under the transformations $u = 2(\ln f)_x$ or $u = 2(\ln f)_{xx}$. Thus, we would like to consider the trial solution is the sum of two linear polynomials as follows:

$$f = g^2 + h^2 + a_9 \text{ where } g(x,y,t) = a_1 x + a_2 y + a_3 t + a_4, \ h(x,y,t) = a_5 x + a_6 y + a_7 t + a_8, \quad (12)$$

where $a_i, 1 \leq i \leq 9,$ are real parameters to be determined.

### 3.1 The Jimbo-Miwa Equation

It is difficult to find the solution of the (3+1) dimensional equation. So, we reduce the dimension setting $x$ or $y$ for $z$. At first, we would like to search lump solution to the Eq. (9) or Eq.(10), replacing $z$ by $x$, where the bilinear equation Eq.(9) is converted to

$$2(3f_{xx}f_{xy} + 2f_{ty}f - 2f_t f_y - 3f_{xx}f + 3f_x^2) = 0. \quad (13)$$

Consider the Eq.(12) as the trial solution similar to the equation Eq.(13).

Inserting Eq.(12) into Eq.(13) and solving for unknown parameters $a_i; (i = 1, 2, \cdots \cdots, 9)$, we obtain a set of constraint results:

$$\left\{ \begin{aligned} &a_1 = a_1,\ a_2 = a_2,\ a_3 = \frac{3\{(a_1^2 - a_5^2)a_2 + 2a_1 a_5 a_6\}}{2(a_2^2 + a_6^2)},\ a_4 = a_4,\ a_5 = a_5,\ a_6 = a_6, \\ &a_7 = -\frac{3\{(a_1^2 - a_5^2)a_6 - 2a_1 a_2 a_5\}}{2(a_2^2 + a_6^2)},\ a_8 = a_8,\ a_9 = \frac{(a_2^2 + a_6^2)(a_1^2 + a_5^2)(a_1 a_2 + a_5 a_6)}{(a_1 a_6 - a_2 a_5)^2} \end{aligned} \right\}$$

The quadratic polynomial function solutions in Eq.(12) with the above set of results, involved six free parameters of $a_1, a_2, a_4, a_5, a_6, a_8$, in turn, yields a class of lump solutions under the determinant condition:

$$\Delta = \begin{vmatrix} a_1 & a_2 \\ a_5 & a_6 \end{vmatrix} \neq 0. \tag{14}$$

and $a_1 a_2 + a_5 a_6 > 0$, \hfill (15)

guarantee the positiveness of the corresponding quadratic function $f$ and the condition

$$(a_1^2 - a_5^2)a_6 - 2a_1 a_2 a_5 \neq 0 \text{ and } (a_1^2 - a_5^2)a_2 + 2a_1 a_5 a_6 \neq 0, \tag{16}$$

guarantee the localization of $u$ in all directions in the $(x, y)$-plane. Now the parameters in the Eq.(12) with the above set of results produces a class of positive quadratic polynomial solutions to the (2+1)-dimensional Jimbo-Miwa like equation Eq.(13):

$$\begin{aligned} f = &\left( a_1 x + a_2 y + \frac{3\{(a_1^2 - a_5^2)a_2 + 2a_1 a_5 a_6\}}{2(a_2^2 + a_6^2)} t + a_4 \right)^2 \\ &+ \left( a_5 x + a_6 y - \frac{3\{(a_1^2 - a_5^2)a_6 - 2a_1 a_2 a_5\}}{2(a_2^2 + a_6^2)} t + a_8 \right)^2 \\ &+ \frac{(a_2^2 + a_6^2)(a_1^2 + a_5^2)(a_1 a_2 + a_5 a_6)}{(a_1 a_6 - a_2 a_5)^2} \end{aligned} \tag{17}$$

and $a_1, a_2, a_4, a_5, a_6, a_8$ are arbitrary constants.

The resulting class of solutions, in turn, yields a class of lump solutions to the (2+1) dimensional Jimbo-Miwa like Eq.(10) under the transformation: $u(x,t) = 2(\ln f)_x$, where $f$ is defined in Eq.(17) satisfying the above conditions Eqs. (14), (15) and (16).

There are six free parameters $a_1, a_2, a_4, a_5, a_6, a_8$ involved in the solution. The solutions are well defined and positive, i.e., if the conditions in Eqs.(14), (15) and (16) are satisfied. The determinant condition Eq.(14) precisely means that two directions $(a_1, a_2)$ and $(a_5, a_6)$ in the $(x, y)$– plane are not parallel, which is essential in forming lump solutions in (2+1)-dimensionals by using a sum involving two squares.

Secondly, we would like to disclose lump solution to standard Jimbo-Miwa like Eq. (9) or Eq.(10), replacing $z$ by $y$, the bilinear equation Eq.(9) can be converted to

$$2(3f_{xx}f_{xy} + 2f_{ty}f - 2f_t f_y - 3f_{xy}f + 3f_x f_y) = 0. \tag{18}$$

To find lump solution of the Eq.(18), we have to consider trial solution of it like to the Eq.(12). Inserting Eq.(12) into Eq.(18) and solving for unknown parameters $a_i; (i = 1, 2, \cdots, 9)$ yields two set of constraining equations for the parameters:

$$\begin{cases} a_1 = -\dfrac{a_5 a_6}{a_2}, \ a_2 = a_2, \ a_3 = -\dfrac{3 a_5 a_6}{2 a_2}, \ a_4 = a_4, \ a_5 = a_5, \ a_6 = a_6, \\ a_7 = \dfrac{3}{2} a_5, \ a_8 = a_8, \ a_9 = a_9 \end{cases} \text{and}$$

$$\begin{cases} a_1 = \pm I a_5, \ a_2 = a_2, \ a_3 = \pm \dfrac{3}{2} I a_5, \ a_4 = a_4, \ a_5 = a_5, \ a_6 = a_6, \\ a_7 = \dfrac{3}{2} a_5, \ a_8 = a_8, \ a_9 = a_9 \end{cases}$$

Thus the solutions are

$$u(x,t) = 2(\ln f)_x, \tag{19}$$

where $f = \left(-\dfrac{a_5 a_6}{a_2} x + a_2 y - \dfrac{3}{2}\dfrac{a_5 a_6}{a_2} t + a_4\right)^2 + \left(a_5 x + a_6 y + \dfrac{3}{2} a_5 t + a_8\right)^2 + a_9$ , $a_2 \neq 0$ and $a_2, a_4, a_5, a_6, a_8, a_9$ are arbitrary constants. The achieved resulting quadratic polynomial function solutions Eq.(19), involving six free parameters of $a_2, a_4, a_5, a_6, a_8, a_9$, in turn, yields a class of lump solutions to the Jimbo-Miwa like equation, under the condition $a_2 \neq 0$. The determinant condition under the condition $a_2 \neq 0$, which guarantees that two directions $(a_1, a_2)$ and $(a_5, a_6)$ in the $(x, y)$–plane are not parallel and is equivalent to $a_5 \neq 0$.

Therefore, the conditions on the parameters $a_2 a_5 \neq 0$, $a_9 > 0$, (20)

will guarantee analyticity and localization of the solutions in Eq.(19) and thus present lump solutions to the (2+1) dimensional Eq.(10).

And $u(x,t) = 2(\ln f)_x$, (21)

where $f = \left(\pm I a_5 x + a_2 y \pm \frac{3}{2} I a_5 t + a_4\right)^2 + \left(a_5 x + a_6 y + \frac{3}{2} a_5 t + a_8\right)^2 + a_9$ and

$a_2, a_4, a_5, a_6, a_8, a_9$ are arbitrary constants. This is a class of complex solution, turned into lump solution under the conditions $a_2 a_5 \neq 0$, $a_9 > 0$, but have no real physical meaning.

Some 3D plots and corresponding density plots of the above achieve solutions are given with the particular choice of the free parameters. Energy distribution depends upon wave height. So, we provide 2D plot to show the energy distribution of the wave i.e., wave height for different values of $y$. In the mean time, we see that when the squares are real valued and $a_9 < 0$, it gives results with shape ellipse or circle type solitonic solutions or singular soliton solution.

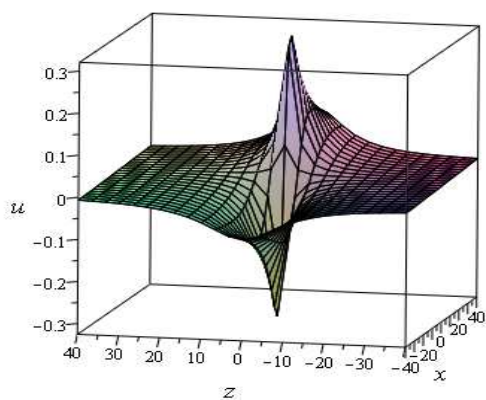
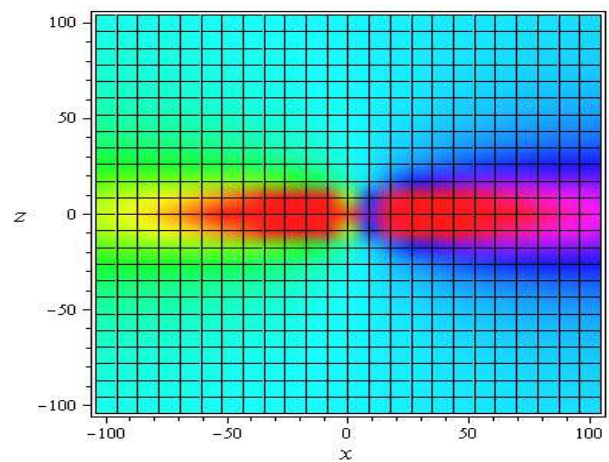

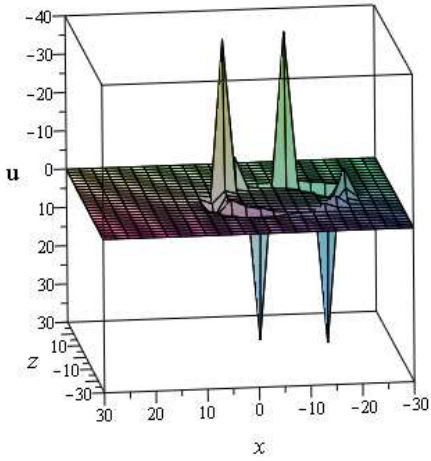
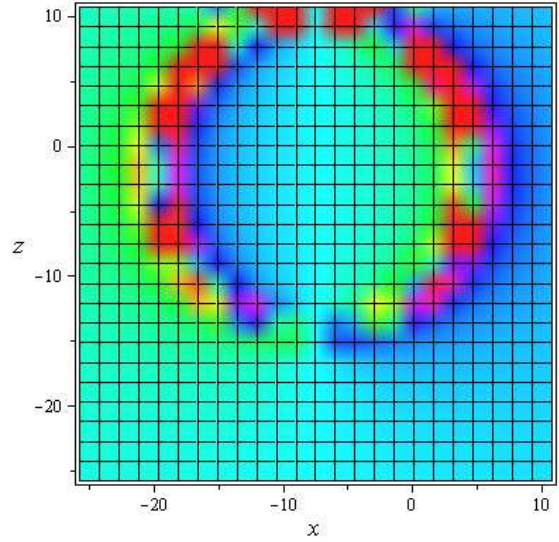

**Fig-1:** Profile of the Eq.(17) (a) for $a_2 = 4, a_5 = a_6 = 1, a_4 = a_8 = 0, a_9 = 40$: **3D plot (left) and 2D plot (right)** at $t = -4$, (b) for $a_2 = 1, a_5 = a_6 = 2, a_4 = a_8 = 1, a_9 = -300$ at $t = 5$.

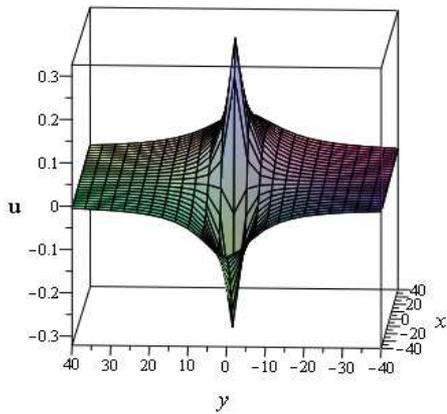

(a)

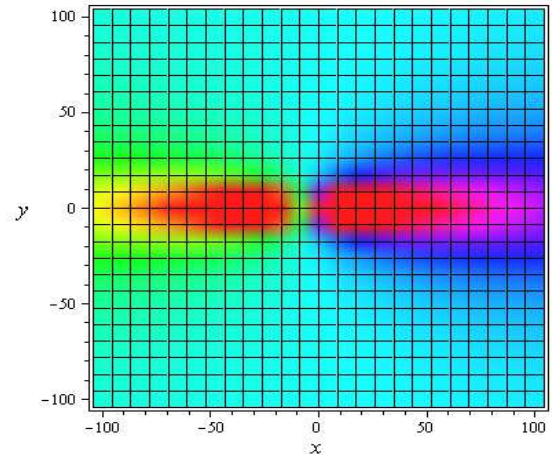

(b)

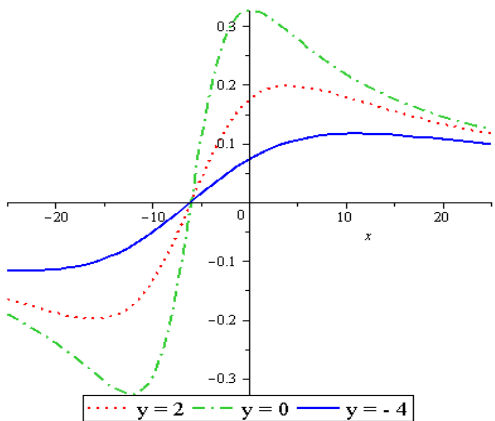

(c)

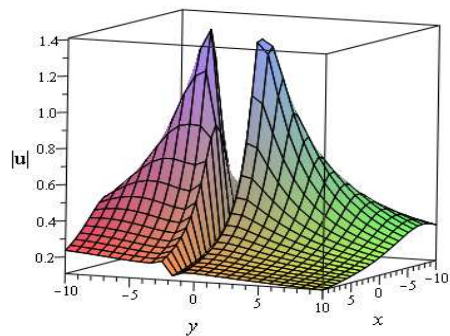

(d)

**Fig-2**: Profile of the Eq.(19) for $a_2 = 4, a_5 = a_6 = 1, a_4 = a_8 = 0, a_9 = 40$: **(a) 3D plot, (b) density plot, (c) 2D plot at** $t = 4$ **and (d) Profile of the Eq.(21) for** $a_2 = a_5 = a_6 = 1, a_4 = a_8 = 2, a_9 = 4$ **at** $t = 1$.

## 4. Concluding remarks and future tasks

Based on generalized bilinear derivatives with prime number $p = 3$, we successfully formulated a Jimbo-Miwa like equation involving a large extent higher order nonlinear terms. We presented positive quadratic polynomial functions solutions to the Jimbo-Miwa like equation. Enough conditions on the exits six free parameters involved in the obtained solutions are proposed to be lump solutions, analytic and localized in all direction of space. Though the proposed Jimbo-Miwa like equation contains more nonlinear terms, their corresponding bilinear form is comparatively smaller and easier to handle than any type of solutions and calculation of these results can be handled very easily with less effort than that of the standard Jimbo-Miwa equation. A complex type lump solution also presented in this article may be useful for exploring new phenomenon in case of complex situations. Circle or ellipse type even simgular soliton solutions are also expressed the achieve solution, when the presented condition are not fully satisfied (generally when $a_9 < 0$). The 3D plots, density plots of the presented solutions with some particular choices of the involved parameters can be found in Figs. 1 and 2 which show energy distribution. Higher order rogue wave solutions could be generated as well in terms of positive polynomial solutions, being a mostly motivating class of exact solutions with rational function amplitudes. Such wave solutions are used to illustrate essential nonlinear wave phenomena in both oceanography [1,2] and nonlinear optics[3,4], which is likely a great deal of current awareness in the mathematical physics society. To survey more soliton phenomena, it would be very remarkable task to consider multi-soliton, multi-component and higher order extensions of lump solutions, more significantly in (3+1)-dimensional cases.